\documentclass[sigconf,balance=false]{acmart}

\AtBeginDocument{%
  }

\setcopyright{acmlicensed}
\copyrightyear{2025}
\acmYear{2025}
\setcopyright{cc}
\setcctype{by}
\acmConference[CHI '25]{CHI Conference on Human Factors in Computing
Systems}{April 26-May 1, 2025}{Yokohama, Japan}
\acmBooktitle{CHI Conference on Human Factors in Computing Systems (CHI
'25), April 26-May 1, 2025, Yokohama,
Japan}\acmDOI{10.1145/3706598.3713317}
\acmISBN{979-8-4007-1394-1/25/04}

\usepackage{graphicx}
\graphicspath{{./images/}}

\begin{document}
\title{Friend or Foe? Navigating and Re-configuring ``Snipers' Alley''}

\author{Andrew C Dwyer}
\affiliation{%
  \department{Information Security Group}
  \institution{Royal Holloway, University of London}
  \city{Egham}
  \country{UK}}
\email{andrew.dwyer@rhul.ac.uk}

\author{Lizzie Coles-Kemp}
\affiliation{%
  \department{Information Security Group}
  \institution{Royal Holloway, University of London}
  \city{Egham}
  \country{UK}}
\email{lizzie.coles-kemp@rhul.ac.uk}

\author{Clara Crivellaro}
\affiliation{%
  \department{Open Lab}
  \institution{Newcastle University}
  \city{Newcastle upon Tyne}
  \country{UK}}
\email{clara.crivellaro@newcastle.ac.uk}

\author{Claude P R Heath}
\affiliation{%
  \department{Department of Media Arts}
  \institution{Royal Holloway, University of London}
  \city{Egham}
  \country{UK}}
\email{claude.heath@rhul.ac.uk}

\begin{abstract}
In a `digital by default’ society, essential services must be accessed online. This opens users to digital deception not only from criminal fraudsters but from a range of actors in a marketised digital economy. 
Using grounded empirical research from northern England, we show how supposedly `trusted' actors, such as governments, (re)produce the insecurities and harms that they seek to prevent. 
Enhanced by a weakening of social institutions amid a drive for efficiency and scale, this has built a constricted, unpredictable digital channel. 
We conceptualise this as a ``snipers' alley''.
Four key snipers articulated by participants' lived experiences are examined: 1) Governments; 2) Business; 3) Criminal Fraudsters; and 4) Friends and Family to explore how snipers are differentially experienced and transfigure through this constricted digital channel.
We discuss strategies to re-configure the alley, and how crafting and adopting opportunity models can enable more equitable forms of security for all.
\end{abstract}



\maketitle

\section{Introduction}
Trust in digital interactions is difficult to gain and easily lost. 
An ecosystem of service providers worry that they may be deceived by their service users or others purporting to be service users.
Whereas service users worry that they may be deceived---or `tricked'---by service providers, others purporting to be service providers, as well as by other service users. 
Meanwhile, essential services key to everyday life, such as housing, welfare, finance, education, and health services are now commonly delivered online as in-person services have been deemed too costly, inefficient, or unable to scale in the contemporary digital economy.
This has generated a `digital by default' digital economy (3DE), where trust has become redistributed, requiring capabilities and support that can disadvantage the most vulnerable in society.

Marketisation has come to dominate how service prov\-iders in the 3DE~\textit{interact} with `customers',~\textit{support} `users' as well as shape the~\textit{values} of contemporary society.
The drive of marketisation to reduce costs, increase efficiency, and scale up, has provided fertile ground for commonplace digital deception---both legitimate and illegitimate---as services have been physically removed from communities, reducing the proximity and social forms of trust between service users and providers.
To address and manage the reduction of social forms of trust, technical validation has become a tool to manage digital deception in the contemporary 3DE, within which both service providers and users alike increasingly perceive and experience the 3DE as (potentially) deceptive and fraudulent.
As (digital) access to essential services is integral to everyday wellbeing, service users must also engage with services using increasingly automated systems and dark patterns (DP), where interactions are adversarial by default, with minimal customer support, and misaligned with the values of trust and relationships of their communities. 
Conventionally, service providers have understood adversaries and their associated harms by conducting threat modelling to design services that offer protection from exploitation. 
However, such models do not recognise the wider security and pressure landscapes that service users today experience in the 3DE. 

Across ongoing and multiple interactions with service providers and other actors online, service users enter what we identify as a ``snipers' alley'', a constricted digital channel that people must navigate: often under pressure and at risk of being harmed by digital deception and the information security practices of service providers. 
This pressure and risk is not experienced equally. 
Some service users that have limited capabilities and access to resources experience the 3DE as incessantly hostile, where they are `picked off' by actors online, resulting in a pressurised form of digital service access that renders them particularly vulnerable to information security controls that seek to protect, as well as marketised practices to benefit, service providers. 
Simply, all actors are potentially deceptive to all actors in the 3DE. 
Service users must navigate digital deception and information security controls to access services online and experience this as a snipers' alley. 
In this constricted space, service providers can morph from providing useful and essential services to harmful snipers according to how providers perceive and treat service users.
In turn, service providers encounter service users as deceptive by default, crafting an alley where service users experience snipers differently according to their background, capabilities, and behaviours.

Drawing on embedded research with communities in northern England, this paper identifies and examines four snipers that service users navigate in the 3DE's constricted digital alley---Governments, Business, Criminal Fraudsters, and Family and Friends. 
In engaging with the (security) needs of communities, we propose three \textit{re-configurations} that can build trust, change relationships with digital deception and information security practices, as well as transform the alley itself.
We call for \textit{opportunity models} to examine how re-configuration can be put into action. 
This means challenging how to (co-)design the 3DE so that digital deception and a sniper's alley---perpetuated through marketisation, DPs, and information security practice---look to secure not only service providers' needs but also service users and communities. 

\section{Background}
To understand how a snipers' alley has formed online, we examine the role of marketisation in the 3DE that has promoted digital deception, DPs, responses to digital deception in information security practice and models, and how trust and fraud has been considered within HCI. 
The digital economy has been broadly described as ``all economic activity reliant on the use of digital inputs, including digital technologies, digital infrastructure, digital services and data. 
It refers to all producers and consumers, including government, that are utilising these digital inputs in their economic activities''~\cite[p.~35]{zekos2022political}. 
This includes digital skills, equipment, and digital goods used in services, platform economies, and more broadly the digitised economy (e.g., industry 4.0 and the algorithmic economy)~\cite[p.~13]{zekos2022political,unctad_digital_2019}. 
Key features include mobility in relation to R\&D, software, algorithms, and users (located anywhere); reliance on tech advances and exploitation of machine learning and big data; and business models leaning towards monopoly and oligopoly.

\subsection{Marketisation in the Digital Economy}
The emergence of the digital economy has produced infrastructures, processes, and applications that generate new forms of circulation and relations between technology and people. 
The technologies that have underpinned the transition have fostered `digital by default' practices for service user engagement by service providers~\cite{Schou2019}, enabling both a scaling and speeding up of circulation.
These new circulations are situated, and have been advocated for, within a neoliberal market logic that prioritises growth and efficiency~\cite{harvey2007brief} that sits at the heart of a promise for more streamlined government and profitable business. 
Within design~\cite{stern2019} and HCI~\cite{Yubo2019}, concerns over marketisation and its impact on notions of care have emerged, primarily through engagement with the French philosopher, Michel Foucault~\cite{foucault_birth_2008}.
These works consider how marketisation introduces forms of financialisation that pervade the design and use of technological systems, and where concerns of care are deprioritised.
Within contemporary neoliberal economies, marketisation has then been defined as a concrete administrative and organisational process that produces intensified competition based on price, whether through government initiatives or private business or a hybridisation of the two~\cite{greer2022}. Its three key features include the commodification of goods, openness, and speed and efficiency, which in turn spurs a sharing economy or platform capitalism~\cite{Langley_Leyshon_2017}. 
In this situation, digital technologies, digitalisation, and HCI design are seen not merely as an outcome of, or response to, neoliberalism but also important contributing factors in its articulation and implementation~\cite{stern2019}.
Indeed, scholars have argued that design and User Experience (UX) can be understood as representing a cutting edge form of neoliberalism~\cite{julier2017}, shaped by the economic imperatives and ends it serves~\cite{zuboff2019}. 

Growing scholarship has highlighted `dark patterns' (DPs) in UX---defined as deceitful design that manipulates, deceives, and manoeuvres ‘users’ into taking actions they otherwise would not take~\cite{Gray2018, Gray2023, Nie2024, Narayanan2020}. 
UX designers use their knowledge of human behaviour to develop deceptive functionalities that are not in the user’s best interests~\cite{Gray2018} and are used by companies to extract profit, harvest data, and limit consumer choice. 
Outlining an ontology of different models,~\citet{Gray2024} show how DPs can include obstruction (e.g., easy to subscribe but difficult to unsubscribe), sneaking (e.g., disguised ads, hidden costs and fees), and interface interferences (e.g., forced continuity). 
Such DPs, as~\citet{Chang} discuss, do ``not rise to the level of crime itself, DPs are used in the furtherance of anti-social behaviour online and in support of fraudulent, even criminal, activity''~\cite{Chang}. 
However, there has been limited attention to the role that both `legitimate' and `illegitimate' DPs collectively play in the 3DE~\cite{Halpern_2016} and how they may perpetuate an ecosystem of low trust and fears over digital deception for service providers and users alike.

\subsection{Security and Threat Models}
Information security has relied on the building of models to secure infrastructures, and in turn to protect data and information.
Conventionally, such models orientate to an explicit technological protection, with an assumption that related social, and relational, forms of human security align with those of the technical~\cite{Coles-Kemp2019, Coles-Kemp2020}. 
HCI has for many years queried the premise of the role of users~\cite{Adams_1999}, where individuals can inhabit multiple roles according to the context in which they live. 
The assumed alignment that protecting technological assets of a service provider (e.g., a government's welfare system) aligns with the needs of service users (e.g., a welfare recipient) is one which has been increasingly questioned~\cite{Coles-Kemp2020, Watson}.

This is clearest in examining threat models. 
In recent years, threat models have become a broader collection of methods to consider how threat may manifest. 
Conventional threat models have been modified by researchers examining intimate partner violence (IPV)~\cite{slupska2021, Slupska, Tseng_2022}, based on the assumption that `better' threat models may aid emancipatory outcomes. 
As Coles-Kemp and Jensen~\cite{Coles-Kemp2019} explore with the case of sociotechnical access of newcomers and refugees in Sweden however, it is not only digital literacy and access to digital services that are required, but the wider range of opportunities for people to engage with services on terms that work for them.

More recently, there has been a significant additional form of security model that does not explicitly align to questions of threat. 
What we term as \textit{associative models} derive from ongoing capabilities from statistical methods to identify `suspicious' populations and communities ~\cite{amoore_securing_2017, foucault_abnormal:_2016, amaro_black_2022}, which offer a method for service providers to manage the increasing distance of service providers from service users. 
Algorithmic processing and machine learning models have enabled the \textit{proactive} identification and detection of security threats to a system, such as through pattern recognition technologies. 
Such models form new associations to reconstitute threat~\cite{Dwyer2023}, which may not be explicitly defined in a service provider's threat model.
For service users, this means that `anomalous' behaviour---which may be normal according to the context and lived experience of that service user---reach a threshold to become an anomaly and therefore appear as potentially deceptive behaviour to service providers.
Like DPs, these algorithmic forms of enabling security can appear hidden and can be difficult for service users to understand and challenge.

\subsection{Trust and Fraud}
The use of information security models to limit digital deception and DPs by service providers therefore transform the trust relationships with service users and how service users in turn interact more broadly in the 3DE.
Regarding HCI, ``the opportunities for direct interaction with service providers and support groups are reduced''~\cite{Coles-Kemp2019}. 
This transformation in the mediation of social relations for service users, whether governments, private businesses, or between friends and family, means that who can be trusted becomes difficult to establish, often requiring technical assurance between participants.
As~\citet{mcsweeney1999security} and~\citet{ROE_2008} argue within security studies, we may understand security as exhibiting two forms~\cite{Coles-Kemp2017}. 
Conventional information security, with its focus on protecting data and its associated infrastructure can often prioritise a service provider's `freedom from' the risk of digital deception (\textit{negative security}) and not prioritise a service user's `freedom to' have access that benefits them (\textit{positive security}). 
Most security controls have an ability to exhibit both aspects of negative and positive forms of security according to the context that they are situated within.
For example, access controls can be understood to limit access, but also enabling a user to gain access to information that empowers them in their daily lives~\cite[p. 467]{Coles-Kemp2017}.
When service users face persistent negative forms of security, this removes agency from services users, as shown by~\citet{Watson} regarding the UK's Universal Credit welfare payment. 

A key threat from digital deception for service providers is fraud, which~\citet[p.~1]{Panicker} consider an ``umbrella term that covers harmful intent, suspicious behavior, and identify misrepresentation''. Within HCI, there has been some limited attention concerning fraud. 
Research has tended to focus on how users identify fraud on online marketplaces~\cite{Sanger} and in financial transactions~\cite{Dev, Story, Freed, Bellini, Latulipe} or how technical controls may be better supported to identify and detect fraud~\cite{Zou}. 
In more nuanced contextualisations of fraud, ~\citet{Kotut} explore how sharing of mobile devices opens up opportunities for fraud to be perpetuated whereas~\citet{Varanasi} consider the complex nature of who commits fraud, including family members of low-income women who engage in crowd work in India.

Within organisations, fraud prevention technologies can be categorised into three kinds, broadly aligning to the security and threat models previously identified. 
That is, 
1) as authentication technologies to prove a user is who they say they are using conventional computer security and threat models; 
2) monitoring technologies to track illicit behaviours, and; 
3) pattern recognition technologies that use thresholds to detect potentially fraudulent behaviour. 
It is the transformed proximity between service providers and service users in a marketised 3DE that information security controls aim to secure and, in so doing, are also intended to mediate a loss of social trust in a 3DE. 
It is in this fuzzy space where there is a misalignment between the security goals of service users (who do not wish to be exposed to digital deception by criminal fraudsters, DPs, and overly negative security ecosystems) and service providers (whose marketised logics perpetuate DPs amid the need for scale and speed through reducing proximity with service users but who are also concerned about fraud) that the conditions for a snipers' alley are made. 

\section{Study Rationale and Motivation}
In late 2018, a researcher from this paper's writing team ran a consultation on digital access in partnership with a social enterprise that provides community housing and welfare services in Yorkshire (northern England). 
The goal of the consultation was part of a programme of action research to better understand the needs of a particular group of digital service users so that the social enterprise was able to re-design its digital education programme to make it more accessible and inclusive. 
The challenges to digital access are many and varied. For this group, there were economic restrictions to gaining access to both network connectivity and computers. 
People in this group also struggled to access the right physical space in which to access services due to cramped housing, leading to a lack of accessible, usable, quiet space in which to make the complex administrative decisions necessary when claiming welfare and related services. 
The group as a whole also had constrained resources both financial, and at times, emotional, making them particularly vulnerable both to becoming victim to online harms and from experiencing accentuated harm as a result of digital deception. 

The consultation took place with 5 male participants aged between 20 and 55. 
All the participants had been unemployed for longer than 12 months and were volunteer workers at the social enterprise. 
All the participants were increasingly dependent on digital access in order to claim welfare payments and access shopping, banking and travel services.
The consultation took 3 hours and addressed the following prompts:
\begin{itemize}
    \item What do you use computers for?
    \item What are computers good for and what are computers bad for?
    \item The one thing I’d like the government to sort out about computers…
    \item How might the IT training room  be changed so it’s used more?
\end{itemize}

The responses were captured on Post-It notes and compiled on a sheet of A2 size paper. 
A wide variety of computer uses were discussed, including: job searches, price comparisons, shopping, ordering takeaways, emails, writing CVs, YouTube, music making, looking up places/shops, making spreadsheets to log uses of the IT room and the computers within it, taking computers apart and reassembling them, looking up recipes, booking cinema tickets, sending and receiving text messages, and creating collages of photos and images. 
Computers were regarded as being good for offering choice---for example, choosing whether to order a takeaway online or going in-person to pick up food. 
Takeaway food was regarded as an important area of choice because some of the participants relied on takeaway food as they lacked the basic infrastructure to cook food. 
Choice about food delivery was important due to a lack of transport options and other challenges with going out and collecting food.
In addition, if the takeaway is ordered, there is an opportunity to leave feedback to large franchises (such as Kentucky Fried Chicken (KFC)); presenting an individual with an opportunity to provide their opinion and develop some agency by having a voice. 
Further opportunities included price comparisons and finding something for less money online, which enables choice as to where things are bought from and, to some extent, at what price. 
Another example was an ability to choose where and when administrative tasks, such as job searches, are conducted. 
All the participants were claiming welfare support---such as Universal Credit---and classified as out of work. In return for the receipt of welfare, all participants had strict and significant requirements related to looking for work and being able to complete the necessary forms online offered important flexibility (cf.~\citet{Watson}). 

The participant group strongly felt that the move to digital services gave them choice in important aspects of their lives. 
However, this choice comes at a cost. 
Online access was considered stressful because there is a lack of trust regarding the material presented by service providers. 
A repeated theme throughout the session was trust---the information on the Internet is  untrustworthy.
Examples were given of fake job adverts, job adverts designed to catch out a job seeker, faulty goods, or goods not arriving as described.
The search engine Google was considered as both something that helped individuals to find information, but also as a mechanism that causes problems because it spreads false information.
The difficulties of working out what was true, and what was false, increased the stress of completing job searches, doing online shopping, and so on. 
Newspapers were often viewed as problematic as Google, with no one news outlet regarded as trusted by everyone.
It was palpable that different participants had varied trusted sources for information and the majority of those trusted sources were community, rather than institutionally, led.  
The engagement with this Yorkshire community presented a metaphorical passage that could be described as a snipers' alley. 
Participant descriptions crafted a fraught channel that people must navigate by deciding whether each interaction is fake or not, and where the financial and emotional impacts of making a wrong decision can be particularly harmful for those with fewer capabilities and resources. 

\section{Research Design}
In April 2023, the research team explored the implications of this potential snipers' alley in more detail with a similar community group in the North East of England. 
As with the pre-study group, the community group members were socially and economically constrained and were accessing digital services with limited network connectivity and access to computers. 
The economic and social resource limitations meant that members of this community group are not only more vulnerable to digital deception, including scams and frauds, but also potentially feel the effects more keenly and take longer to recover. 
Digital access was in large part forced upon members of this community group because essential services have moved online, retail and banking infrastructure has moved out of the local area, and local authority support had also moved online.

This research was developed in consultation with the community group and was chosen because, like its Yorkshire counterpart in northern England, it offers welfare and housing support. 
The community group was interested in working on the theme of digital deception because individuals from lower socioeconomic groups were becoming increasingly vulnerable to scams and frauds due to the effects of climbing price inflation that made digital access increasingly pressurised, known in the popular media as the `cost of living crisis'.
The participants were recruited through the community group, and similar to the pre-study participants in Yorkshire, all accessed digital services through necessity rather than choice. 
Participants were self-selecting as they had an interest in fraud. As this research was founded in co-design, we did not impose strict selection criteria in addition to their interest in fraud, apart from one exception.
Both the researchers and the participating community group acknowledged that running a training session on fraud had the potential to transfer knowledge to would-be fraudsters and scammers. 
Participants were therefore carefully selected so that they were more likely to be interested in preventing rather than committing fraud and scams, drawing on the experience of the community workers.
In total,  20 participants were involved the in study from North East England.
In conjunction with community workers, the following research questions were agreed:
\begin{itemize}
    \item What are the digitally-delivered scams and frauds that voluntary and third sector organisations are currently most concerned about? 
    \item How do voluntary and third sector organisations communicate digitally-delivered scams and frauds to the communities they serve? 
    \item How do communities respond to the awareness and training messages delivered by voluntary and third sector organisations on the topic of digitally-delivered scams and frauds?
\end{itemize}

This set of questions led to the formation of three sessions aligned with the three research questions. After the brief was agreed with the community workers, they began recruiting participants for each session: 
\begin{itemize}
    \item \textbf{Session 1}: From a range of voluntary and third sector organisations as well as local government
    \item \textbf{Session 2}: Members of the community group
    \item \textbf{Session 3}: A roundtable of some participants from sessions 1 and 2
\end{itemize}
Each session was granted ethical approval by the researchers' academic institution, with participant recruitment and the subsequent session carried out following the relevant ethics protocols. The data was reviewed at the end of each session and any data identifying individuals was redacted. Ethical approval was applied and granted for each session because the ethical dimensions changed with each session. The first session required a generalised reporting of frauds and scams. The second session contained an element of training and we wanted to avoid training people to commit rather than prevent scams and frauds. The third session meanwhile focused on the personal experience of scams and frauds. 

\begin{figure*}[ht]
    \Description[]{A collage of nine photos in a row and nine digitally created black boxes below each photo. Each photo contains an A2 sheet of paper, where participants have affixed several yellow Post-It notes corresponding to a question at written at the top of each page. The digitally created black boxes are from a corresponding Miro digital whiteboard with yellow notes which transcribe the writing on the Post-It notes from the corresponding whiteboard above.}
    \caption{Post-It notes arranged on A2 paper from Session 2 (top) with their digital equivalent on a Miro board (bottom).}
    \centering
    \includegraphics[width=\textwidth]{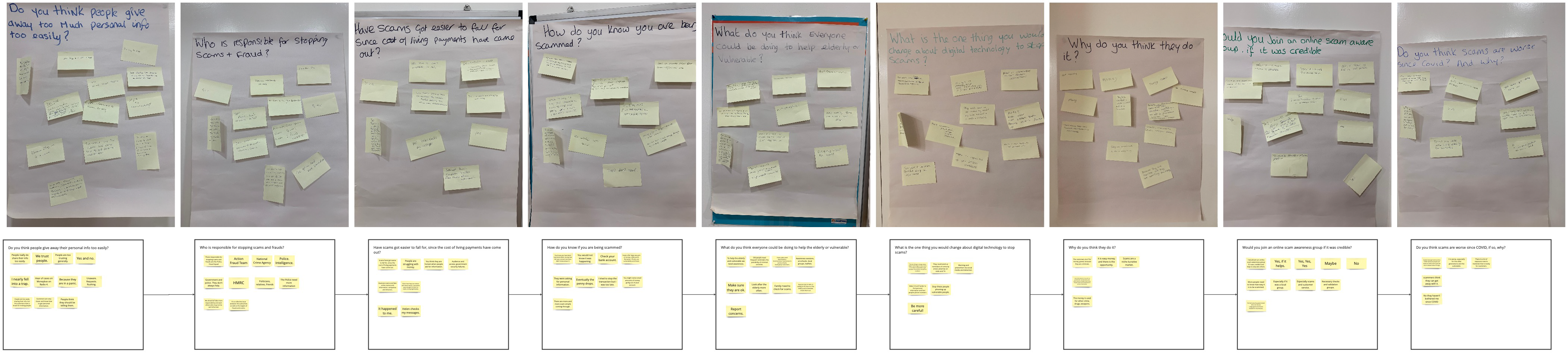}
    \label{fig:Session2_PostIts}
\end{figure*}

Data was gathered from notes made by researchers during and after the sessions as well as notes made by participants. 
Post-It notes were used as the mechanism for participants to provide input. 
Post-It notes are useful because they provide a small space for input, which is less intimidating for participants experiencing barriers to group participation. 
They also have the added advantage of creating a visualisation that shows growing participation as each new Post-It note demonstrates that the volume of feedback is growing. Equally, they can be removed, rewritten, and edited as well as moved and reassembled to build new associations (see~\citet{christensen2019sticky}). 
The Post-It notes were grouped onto A2 paper in each session (for example, see Figure~\ref{fig:Session2_PostIts} for outputs of Session 2), where photographs of the data were taken at the end of each session and uploaded to a Miro board (a collaborative virtual `whiteboard') and digitally transcribed.

\subsection{Session Design}
\label{Section Design}
All three sessions were in the style of community consultations. These sessions were designed to be informal, participatory focus groups using simple worksheets to support discussion.  
Each session lasted circa two hours. 
After each session, notes were shared and discussed with the participant group over Zoom to develop a consensus around the findings and to follow through on the consultation approach. 
Each session was structured in a broadly similar way, with an introduction that set out the background and purpose of the session. 
The questions used in the second and third sessions were formulated using the outputs from the previous session. 

\paragraph{Session 1:}
\begin{figure}[t]
    \centering
    \Description[]{A photograph of A3 white paper sheet, with a range of black boxes with various questions inside. On the sheet, there are seven boxes with a variety of Post-It notes affixed to the sheet of paper. These Post-Its are of three colours (blue, pink, and orange). On the Post-It notes are handwritten comments from participants corresponding to a question within each box on the sheet of paper.}
    \includegraphics[width=\linewidth]{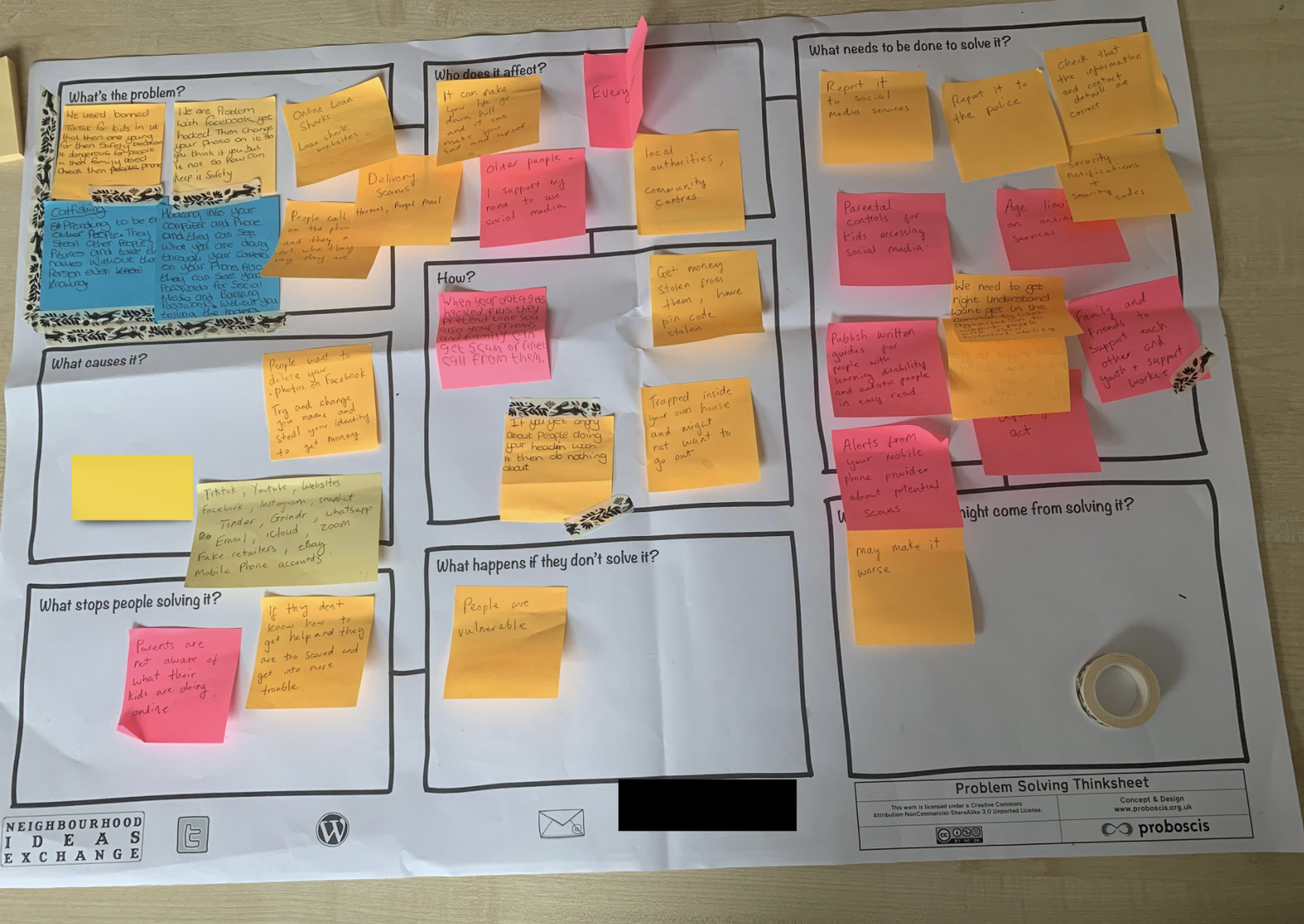}
    \caption{A photograph of a worksheet from Session 1 with Post-It notes.}
    \label{fig:Worksheet}
\end{figure}
For the first session, a broad mix of community actors were recruited from voluntary, third sector, and local government organisations to gain up to date information on current digitally mediated deception on frauds and scams, preferred interventions, barriers and challenges to successful interventions, and the role that privacy and digital identity technologies might play in overcoming these barriers and challenges. 
The participants were divided into two groups. 
The first group was a consultation between community workers, a local social enterprise developing digital services with local community groups, and two representatives from a community liaison team in the local government. 
The second group was a consultation between members of a grassroots local community group supporting those with physical and cognitive disabilities and their carers.
A simple worksheet (see Figure~\ref{fig:Worksheet}) was made available that asked participant groups to reflect on the problem and potential responses.
The groups were encouraged to think about specific types of digital fraud that they encountered and work through each type of fraud using questions presented on the worksheet.

\paragraph{Session 2:}
For the second session, participants were recruited from the community group. 12 participants were recruited; 5 female and 7 male participants in an age range between 18 and 72.   
The community workers set out a series of questions based on outputs of Session 1 that emerged from a post-session review.
The questions were set out on flip chart sheets and participants were encouraged to walk around the room and use Post-It notes to capture their responses to these questions (see Figure~\ref{fig:Session2_PostIts}). 
The questions were as follows:
\begin{itemize}
    \item Do you think people give their information away too easily?
    \item Who is responsible for stopping scams and frauds?
    \item Have scams got easier to fall for and has it become worse since COVID?
    \item Why do you think scammers do it?
    \item How do you know if you are being scammed? 
    \item What do you think everyone could be doing to help the elderly and the vulnerable? 
    \item What's the one change you would make to digital technology to stop scams and frauds?
    \item Would you join an online scam awareness group if it was credible?
\end{itemize}

The facilitator explained the purpose of the session, the role of the researchers, and worked through the participant consent form.
The community training pack included descriptions of everyday frauds and scams that their community experiences.
Participants then formed small groups to discuss work through the same worksheet given to participants in Session 1.
Participants were also encouraged to discuss and note down important tips and skills advice necessary to help people protect themselves against these frauds and scams. 
Taking onboard the observation about stigma and shame around reporting being a victim of scams and fraud, the session was designed as a safe space~\cite{Duarte_Safe_Spaces}, so that participants could talk about personal experiences with digital deception.

\paragraph{Session 3:}
The third session was a roundtable discussion with a focus on developing possible interventions for the issues and challenges identified in the first two sessions. The participants worked as one large group reflecting on the outputs from the previous two sessions and debated what interventions could be made by the community group (both as individuals and as a group) to reduce both vulnerability  and harms to individuals if they are a victim of digital deception. 

\subsection{Data Analysis}
Research materials from sessions were co-analysed and iterated between researchers and participants.
The Miro board became a means of sharing, reorganising, and discussing data with participants.  
The Post-It notes produced from each session were digitised by the research team, with colours of each group being maintained in order to maintain consistency across the data analysis. 
Colour-coding enabled a tracing of ideas as the data was increasingly re-located among other materials.
The process of iterating upon the previous session's data created a layering and collaging of discussions, thoughts, and ideas~\cite{Heath2022}, allowing a malleability and reordering both in-person and afterwards digitally.
Here, the distinction for both participants and researchers between generation and analysis of research materials was non-linear and challenged throughout the study.

After the sessions concluded, materials were rearranged on Miro by another of the paper's authors to provide a schematic mapping of the discussions across the three sessions. 
This mapping enabled connections between `illegitimate' actors and activities, including illegal scams and frauds, with `legitimate' actors and activities. 
This schematic offered an insight into the entangled relations between scams, frauds, legitimate actors, and people involved in the study.
This schematic mapping was subsequently used to conduct reflexive thematic analysis among the research team (drawing on the discussion of~\citet{Braun_2021}), alongside researcher notes. 
This enabled connections to be drawn between the particular contextual arrangements of this community in North East England. 

In response to the reviews of this paper, in November 2024 we returned to the community group to present our re-evaluation of our findings and for the group to reflect and contribute to the themes we re-designed. Two of the paper's writing team returned to the community group, with one author engaging on fraud and the new themes with the community group whilst taking hand-written notes for around two hours (after having a collective hot lunch and drinks with the community group and community workers to catch-up informally). 

\subsection{Limitations}
This study builds on a long-term substantive engagement with one community in North East England, based on a pre-study of a community group in Yorkshire, northern England. 
Our study emerges across and through deep, and sustained, interaction with this community over the past 10 years. 
This paper presents findings based on co-design and trust that has been developed with community workers and the community in this time.
We argue that this study is not easily replicable as much as the context, experience, and depth of our relationships with these communities offer embedded insights into digital deception in the 3DE for HCI, co-design, and communities themselves with wider applicability.
We acknowledge that our sample size is notionally small. 
However, for co-design, we proffer, that deep engagement offers opportunities for transferability of methods and themes we present next.
We contend that although this community's engagement on fraud is not generalisable---in that the particular contexts and histories of northern England (re)shape how snipers' alley is felt, experienced, and has impact---many other people and communities have limited capabilities and resources that may not be immediately categorised as `vulnerable' but are worthy of further research through embedded practice and co-design.
Despite the limitations of (scientific) sample size and generalisability, we believe that if using different mechanisms at a larger scale, it would be difficult to gain trusted insights on what can be a very difficult, deeply-personal, and sometimes ``embarrassing'' topic. 

\section{Findings}
In this section, we outline our findings across three key themes that emerged from our analysis: the role that \textit{blurred lines} play in the 3DE between actors, the role that \textit{systemic stress} plays for people navigating snipers' alley, and how \textit{capacity} is central to the agency of individuals and communities within the 3DE.

On our return to North East England, participants reflected on the snipers' alley metaphor.
During the discussion, one participant outstretched their arms, articulating a virtual manifestation of the alley, stating that one can ``walk down and [they] take everything''. 
This taking of everything by an implicit `they' was discussed by the group and included not only criminal fraudsters, but also the vast amounts of data being collected about them by service providers and a sense that they ``get pestered'' to do things that are not in their best interest (such as from DPs). 
In reflecting on their engagement with the 3DE, another participant clearly affirmed that without such digital services, ``I'd have nowt''\footnote{In the dialect of northern England, this is equivalent to ``I would have nothing''.}.
It is this requirement to engage with the 3DE that made the metaphor of snipers' alley so visceral for the community group on our return---they could visualise navigating the alley and how the actors in the 3DE morph and cohere in ways that make digital access fraught with potential digital deception.

\subsection{Blurred Lines}
For service users navigating sniper's alley, there are multiple points where lines are blurred. This is caused by a `flattening' of social relations in digital interactions, where actors and `legitimate' processes can often be confused, resulting in a 3DE that is considered low-trust and deceptive.

Across all sessions, it was clear that there is a lack of trust in \textit{who} or \textit{what} service users are engaging with online due to the decrease in proximity instantiated through a movement to digital-only provision of essential services.
This had been amplified by the withdrawal of in-person services of businesses, including the closure of banks and post offices, with one reflecting that this had resulted in fewer services in the community as ``in the town, we've got no shops!''.
This wider effect of the 3DE means that many places where service users could engage in-person; whether for government assistance, retail, or financial services, many had been removed.
For service users who may be at risk of digital deception, community anchors---individuals working in service provision---have often gone, where no one person has a cohesive view of people in the community. In the past, a trusted local shopkeeper or post master may have identified a fraud or scam through knowing a person and unusual activity.
However, these social relations have now significantly reduced.
This removal of in-person services and their provision in the 3DE mean that engagement has become `flattened' with service users being forced to access services on digital devices.
Rather than trust built up in place with communities, service providers in the 3DE instead used technical validation to secure their interests.

The lack of capability to socially trust actors and entities online is what facilitates the capacity for scams and fraud.
Participants discussed the difference between fraud (when someone pretends to either be someone they are not, or represent a trusted entity when they do not) and scams (when someone tricks an individual into doing something that harms them, such as parting with money). 
Participants noted that they feel that they can be `tricked'~\footnote{In the dialects of North East England, `trick' does not simply refer to being tricked, or the act of a trick, but also being gullible or that someone is out to `get you'.} into parting with money or giving information such as passwords, bank account numbers, as well as personal and sensitive information by both illegitimate and legitimate actors, blurring the line between digital deception and criminal activity.
This was particularly pertinent to the participants, where many had noted that they had been a victim of fraud or knew someone who had been. 
As one participant reflected, a criminal fraudster had scammed them out of money through an urgent request from a friend's `hacked' account on Facebook, using its Messenger function.
Due to their reliance on the social trust of a friendship, this participant had sent money to a criminal fraudster.
Meanwhile, they later explained how they had received a letter demanding payment from their energy provider, \textit{British Gas}.
This letter appeared fraudulent and they did not trust the phone number on the letter to ring the service provider.
When they did contact the provider by phone from another letter, it was noted that this was an erroneous payment demand.
Although this may not be an example directly derived from a digital interaction, the community group was clear that the 3DE had caused a drop in trust with \textit{all forms of interaction with service providers}.
Here, the everyday lived experience of participants mean that even conventional distinctions between actors and processes have become exceptionally blurred.

These complex interactions between multiple actors---in this case criminal fraudsters, a service provider, and friends and family---can appear to a service user as both potentially caring and malicious in different contexts. 
It can also happen in the intersections between online and offline experiences.
As much as participants reflected that they might not trust anyone, they remained reliant on their family and friends and the broader community for support, even if some noted that the former can (un)intentionally deceive them.
Some participants try to resolve the blurred lines of the 3DE by relying on the community group's IT resources, with one stating ``I only do online activities at [the community group]''.
For service users, the increasing distance from service providers in the 3DE means that all actors become interconnected and `flattened' when they interact.
That is, service users find it difficult to identify who and what they are engaging with, and whether they are able to trust their interactions, blurring the line on who to trust and not.
Participants also identified a wider range of problems: including the security and influence of other social media companies such as TikTok, catfishers (entities pretending to be someone they are not) and online financial `shark' companies~\cite{Ash} as examples of actors that crossed the blurred lines of digital deception. 

\subsection{Systemic Stress} 
The marketised logics of the 3DE, combined with scams and frauds, make access to even essential services systemically stressful according to our participants. 
Not only do service users have ``a lot to get used to'' in terms of processes, interfaces, and security systems, but these are often combined with temporal pressures, multiple communication channels, and increasing isolation.

Many interactions with service providers occur in short bursts, often requiring the return of information or signing up for a deal within tight time scales.
For example, there may be requests for information for Universal Credit welfare payments required suddenly, with the risk that not providing the information will lead to a sanction (deduction of welfare payments). 
This is often replicated by criminal fraudsters who require information quickly and under pressure of a loss or missed opportunity.
In addition to the quick responses required, interactions are frequently required at pressurised times of the year, such as at Christmas.
At such key moments, stress is amplified for participants. 
Participants reported that scam messages appeared more often, meaning service users are exploited when already stressed with other social and fiscal pressures.
The perceived power imbalance between service users and government, alongside stigmatisation and embarrassment of service users to ask for help or assistance, was identified as a key reason why it was difficult to run awareness campaigns as well as encouraging individuals and groups to talk about their fraud experiences.
As a result, people were not claiming tax rebates, additional welfare entitlements, or engaging with government anti-fraud messages. 

At a time where `cost of living' and heat payments were being managed by different government actors, stress was exacerbated by multiple channels for communications---from SMS, email, to postal letters.
As in the case of the participant who explained their interaction with an erroneous payment demand from an energy provider in the previous section, there is a pressure to engage with multiple sources of information in different ways.
To access information online often requires multi-factor authentication, which is exploited by criminal fraudsters to deceive service users (and service providers).
A lack of a single trusted communication channel makes navigating the 3DE for service users increasingly pressured to evaluate multiple---often conflicting---sources of information.

The move away from in-person service provision was most keenly felt through isolation.
Participants reflected that without the community group or family and friends, they would find it difficult to assess digital deception.
However, in a marketised 3DE, such isolation enables the sale of goods.
As some participants often did not have others to rely on at all times, no one was ``looking out for them'', leading them to take actions that they later reflected on as being deceptive by (legitimate) service providers. 
Participants often feel overloaded with everyday life and this made them feel more prone to digital deception.
These broader pressurised environments shape how service users engage with digital services, and view them as adversarial to achieving their other security goals (e.g., being able to receive welfare payments to live).
These created systemic stresses---where DPs and information security practices, including authentication---make accessing the 3DE tricky and potentially deceptive as they navigate snipers' alley.

\subsection{Capabilities} 
Each session, and in particular the third, reflected on the capabilities that various actors, including participants, have and how they may wish to respond to digital deception. 
The responses to the question ``Who was responsible to limit digital deception?'' still heavily relied on the state.
A common sentiment was that ``they're better equipped than we are'' but also that the nature of criminal fraudsters meant that police forces find it difficult to track criminal fraudsters down. 
Likewise, there was a role for service provides. 
One instance discussed the role of banks in preventing fraud by monitoring bank account activity, in what was described by one participant as ``a good type of snooping''.
Another example included telecommunications providers automatically blocking scam callers.
Participants however considered criminal fraudsters ``very clever people'' that ``mostly come from abroad''. 
A persistence that criminal fraudsters are clever, and therefore difficult to avoid, permeated many of the conversations. 

Overwhelmingly, the group felt that government, police and fraud protection initiatives were responsible for stopping scams and frauds.
The group also argued that everyone should take a collective responsibility to resist scams and frauds.
The group argued that the more economically vulnerable people become, the more vulnerable they are to scams. 
This was complemented by how to help vulnerable people, with awareness, family, and `check-ins' by the community or social workers being important.
The group reflected that scams and frauds are not easy to spot and it can take time to know that you have been tricked.
Awareness, collective responsibility, and increased regulation were all regarded as important mitigations. 
Information security mitigation included privacy and confidentiality controls.
The group also advocated for better warning technology.
At the same time, the participants were keen to join a group focused on online scams to build up skills, knowledge, and confidence so that the group \textit{could take action}. 

When accessing the 3DE, service users are often understood to hold certain capabilities within security models. 
Participants were clear that they required capabilities to be able to respond in their own way.
However, services providers should be genuinely ``listening'' to their concerns, and communities being offered social funds to help.
What is clear is that this group was highly engaged and had developed awareness of frauds and scams.
Participants concluded that digital skill building, accessible communication, education and training, and encouraging people to talk about their experience of being scammed and defrauded were all important interventions. 
However, isolated and individualised responses were not enough. 
As one participant noted, training and awareness is ``good for some, not for others''. 
This was supported by the collective group. 

If community interventions were to be successful, participants argued that it was important to be focused on specific scams and frauds rather than digital deception more widely. 
In session three, the following initial list of scams and fraud was created that were felt to be of most interest to local people:
\begin{itemize}
    \item Heating scams and frauds---the focus of these attacks are to exploit the emergency heating payments, either by tricking an individual into handing over the heating payment or fraudulently claiming to be the government offering a heating payment in a bid to gain bank details.
    \item Romance scams---the focus of these attacks is for a fraudster to pose as an individual with a romantic interest in the victim in an attempt to elicit money from the victim.
    \item Banking scams---the focus of these attacks is to get access to banking details.
    \item QR code scams (Quishing)---these attacks have become prevalent since QR codes became prominent during the COVID-19 pandemic. The attacks re-direct victims to fake websites or to websites with malware in a bid to gain access to personal details (in particular banking details). 
\end{itemize}

The group identified four conditions that were necessary in order for community groups to successfully take action to combat fraud and scams:
1) The community itself must want to take action; 
2) The community group must know that they can make a difference;
3) The community must be confident to take action, and; 
4) The community group must be knowledgeable about frauds and scams.
The participants identified that it was not enough to have awareness about frauds and scams but an individual must be confident that they can take action to reduce their vulnerability to digital deception. 
Specifically, this means being able to take action if a person becomes victim to a scam or subject to an attempted scam. 
It also means increasing awareness and practical knowledge about how to keep personal details private. 

Participants concluded that a community action group could help individuals to gain confidence in taking action by making individuals aware of which advice can be trusted and which cannot. 
As part of this trust building process, it was also felt that community groups could work to debunk conspiracy theories so that trust was built on facts and tested knowledge. 
They also felt that a community group should work to help individuals resist emotional blackmail from fraudsters and scammers as well as `legitimate' deceptive practices (including DPs). 
Such a community group should also work to develop a route to contacting vulnerable members of the community. 
Whilst there is an emphasis on awareness, the following digital design points were identified:

\begin{itemize}
    \item Interventions that change the trade-off between the speed and security of interactions.
    \item Interventions that change the ways that customer service is undertaken. This includes re-design to focus on the empowerment of the individual seeking help by using messaging in easy or simplified English, or by re-designing customer service to focus on care rather than efficiency.
    \item Interventions that build a route to contacting vulnerable members of the community.
\end{itemize}

These interventions are not only points at which technology can be used to support and protect the individual using a service but can also be regarded as re-configurings of digital interaction.
This is in order to both reduce opportunities for frauds and scams and to develop resilience to digital deception more broadly.

\section{The Snipers}
Across the study, four key actors emerged as those who could conduct digital deception against people, both legal and illegal, legitimate and illegitimate. 
They are fundamentally blurred in the 3DE.
The first two could be considered as service providers: 1) Government and 2) Business. 
Then there are illegal actors 3) Criminal Fraudsters whose aim is to explicitly scam and defraud people and 4) Family and Friends who may seek to exploit the trusted relations between people. 
These actors can be characterised as snipers, as they often trick, can be hidden, are sometimes trusted, but can cause harm if a person's capabilities to respond are limited or they are more susceptible to being `tricked'. 
All people must pass through the constricted alley of the 3DE, but the alley's actors transfigure according to the capabilities, resources, and behaviours of those who pass through. 
Those with greater capabilities---whether it is more education, awareness, finance, support, and so on---can better manage interactions with actors because their `social envelope' and resilience mechanisms provide a capability to become either invisible or `normal' to actors. If they become visible or anomalous, they are resistant, or able to absorb implications of interaction with an actor turned sniper.
For some people, with more capabilities, the snipers may not appear as a threat---as they may have good technical know-how, have social back-ups, or they may only infrequently engage with actors (such as governments). 
Indeed, they may be more akin to a `normal' user, passing monitoring and pattern recognition checks for fraud and scams.
Those people without such protections become exposed to actors that are transfigured as snipers, where the constricted digital alley becomes increasingly difficult to navigate and can be harmful.
The marketisation of the 3DE and security models have accentuated the constricted nature of the alley, by enabling a `flattening' of social relations through moving services online and removing these from communities in-person.

Participants described when carrying out online transactions that they did not know if they were to be supported or tricked by service providers. 
This sense of insecurity, not knowing whom or what you can trust, creates a sense of being ``picked off'' as an individual transacts or takes part in digital interactions. 
Each message, each request for information, each call to action may (or may not) be what it seems with very little information, know-how or capability to make the decision to trust or not trust. 
Given that our participants---and many service users---must use digital services, as this is the dominant mode of engagement or because there is no viable alternative, the sense of being forced to run the risk of being tricked and harmed, has increased. 
The snipers embody the blurred lines, systemic stress and the capability gaps we identified in our findings, which in turn provides the ground for an alternative means of modelling the service user experience of being online and under threat in the 3DE. 

\subsection{Sniper 1 - Governments}
The majority of our participants were claiming state welfare or pension from the government. 
Over the last decade, this type of interaction has become digitised. This made some of our participants vulnerable to either under-claiming or over-claiming welfare. 
The community workers reported cases of individuals and families under-claiming for fear of being aggressively penalised for claiming too much money. 
The threat of ``getting it wrong'' was reduced for our participants by working with the hosting community centre, who supported digital interactions with the government, explained the nature of the interactions, and what data was being collected from the claimant and why. 
Participants provided examples of where they felt digital interaction with the government was adversarial and instances where the government could transfigure from being a supportive actor or carer to a sniper. 
An example is when applying for the Universal Credit welfare entitlement~\cite{Watson}. 
To apply online---as it is predominantly an online service---requires a smartphone with a camera, an internet connection, and identity documentation. 
The application process for Universal Credit requires significant experience of online know-how, and to maintain this welfare support, there are frequent digital messages with the government. 
If certain conditions are not met (e.g., filing an online journal), then entitlements may cease. 
This pressure to ensure digital connectivity and know-how, as well as pass security checks at a distance, make government support appear more like a sniper to those with fewer capabilities. 

\subsection{Sniper 2 - Business}
Our participants gave many examples of non-government service providers, including telecommunications providers, train operators, and travel agents where they felt they had been tricked into buying a service that did not meet their expectations or where exiting from an arrangement proved extremely difficult. 
The participants commented that often, the service providers used similar tactics to the criminal fraudsters---such as dark patterns (DPs)---but law and regulation meant that they were being legally manipulated to part with their money. 
Our participants pointed out that this made it harder to tell the difference between a criminal fraudster and legitimate business. 
For example, terms and conditions that do not use ``easy English'' or systems that require divulging personal information to access, blur the boundaries of what is considered potentially fraudulent and not.

\subsection{Sniper 3 - Criminal Fraudsters}
A combination of inflation of food and other essentials, the increased cost of utilities such as gas and electricity, together with increases to rent and mortgages, created challenging conditions for our participants. 
Between 2022 and 2024, the UK government provided additional `cost of living' payments to those in lower socioeconomic groups. 
However, our participants gave examples of where either they or their family and friends were reluctant to respond to so-called ``crisis payment'' messages for fear of being scammed by criminals. 
This fear is grounded in examples of where such messages have turned out to be fraudulent and previous experience of fraudsters taking advantage of non-routine payments and ad-hoc welfare schemes where eligible claimants are not quite sure how the scheme should work.
In particular, the growth of more complex eligibility requirements and `targetted support' has enhanced this problem.
Criminals took advantage to pretend to be a supportive service provider when, in fact, they were a sniper seeking to exploit the uncertainty of engaging with government. 

\subsection{Sniper 4 - Friends and Family}
Friends and family were often a part of our participants' digital interactions. This was either because they were aided and supported by family and friends in completing digital interactions or because friends and family were part of their digital interaction community (e.g., via Facebook).
Participants gave examples of where friends and family were either an unknowing conduit for frauds and scams or where friends and family abused their position of trust, shifting from supporter and enabler to sniper and back again. This type of sniper is well-researched within the IPV literature~\cite{slupska2021}. Such family and friends remain ambiguous for addressing frauds and scams and broader digital deception, as they may be trusted at times, but not at others. 

\section{Discussion - Navigating Snipers' Alley}
Our findings provide an insight into how digital interactions in the marketised 3DE, combined with information security models and practice, aided deceptive digital practices that produce uneven experiences for service users for whom digital access is a necessity rather than a choice. 
The findings also show how digital protection mechanisms, legislation, and awareness of frauds and scams fall short of providing the support that service users need.  
Our findings demonstrate how digital deception and mediated interactions under current socio-economic conditions do not simply cause harms to people and communities, but more profoundly reshape and redefine \textit{relations} at all levels of contemporary societies, where people do not know what and how they can trust, and where family, business, and government actors can suddenly morph into snipers. 

In this section, we look at three key ways that HCI can work with grassroots groups, voluntary and third sector organisations to both improve, transform, and centre service users within the 3DE. 
These are: re-configurations of digital \textit{interaction}, re-configurations of \textit{support}, and re-configurations of \textit{values} that can challenge a marketised 3DE. 
Central to these three approaches is the recognition of the four potential snipers that works across blurred lines, systemic pressures and gaps in capabilities.
By exploring the transfiguration of actors into snipers, we offer an alternative approach to modelling service design. 

\subsection{Re-configuration}
Across the study, participants made several suggestions to reduce the potential for successful digital deception, including directions for re-designing the \textit{interaction}, \textit{support}, and \textit{values} of digital services in the 3DE. 
Here, we build on their suggestions to develop proposals for the HCI and security and privacy communities to make digital users not only less vulnerable to digital trickery and more resilient to the harms that might arise, but to be able to re-configure how they engage with service providers so both the snipers and the alley no longer exist in the way that they are currently articulated in this paper. 
Each of these re-configurations can operate on differing temporal scales: short, medium and long term.  These proposals combine to form the basis of an opportunity modelling grounded in positive forms of security which both challenges and, potentially complements, traditional forms of threat modelling grounded in negative security. 

\subsubsection{Re-configuring Interaction:} 
Participants articulated online experiences characterised by pressures and anxieties stemming from an \textit{ecosystem} of digital interactions, including the speed of interaction and isolation from support---no matter \textit{who} the sniper might be. In our findings, we characterised these as systemic stresses and our findings show that snipers gain more power when security practices are rushed.

Our participants noted how vulnerability was compounded by a range of digital deceptive practices and actors exploiting stressful times of the year, for example at Christmas. 
HCI has, for example, explored visual counter-measures for DPs in interface design~\cite{Schafer_2023}, noting that different applications of DPs require alternative responses also in line with users preferences, digital capabilities, and needs. 
Building on this work, our insights show the need to take an embedded view of the ecosystem to more wholly understand and appreciate the experience of service users, and critically consider the temporal and contextual dimensions of \textit{when} interactions take place, so as to develop responses and countermeasure responsive to people and communities. 
Our study points to the need to engage more critically with the trade-offs at play between the speed of interaction and security---whereby privileging the first can in practice significantly diminish the latter. 
The temporal dimension in digital deception has been noted in the context of DPs in commercial services~\cite{Gunawan_2022} and gaming temporally-oriented interactions to ``cheat'' users~\cite{Zagal1043332}. 
Slowing down interaction might be seen as counter to current trends in a marketised 3DE---where constraining the speed through which users can perform certain tasks can be understood as a method to identify potential frauds and scams, and where this is not possible, to permit automation through pattern recognition to identify anomalous behaviour. 
Taken together, critically examining how increasing the cognitive load of service users already under pressure, as well as evaluating how authentication, monitoring, and pattern recognition technologies restrict access, is required in order to understand how people are excluded and thus perpetuate \textit{negative security} across the 3DE.

\subsubsection{Re-configuring Support:} 
Another avenue for intervention concerns the re-configuration of the channels and modalities of communication between service providers and users. 
For the study's participants, such re-configuration needed to focus on the empowerment of individuals seeking help through the use of communication channels and messaging that were easy to understand and used simplified English. 
By making the messaging useful and accessible, it is possible to reduce the power of the snipers and create a greater sense of agency in service users to engage with digital services. 

Most critically, participants suggested that so called ``customer service'' could be re-configured around notions of ``customer care'' that centres practices of care over marketised logics of efficiency, speed, and scale. 
Care for our participants was helpful in highlighting the relations of mutual obligation and responsibility for the well-being of people over time.
Such care would help to reduce the likelihood of the accidental creation of snipers and also to create greater capacity to respond to snipers as they emerge.

Services' duty of care was highlighted by participants, and needed to be enacted by creating the conditions for people to understand routes available to them, and critically that there was trust that something would be done as a result of their action (i.e., efficacy). 
In HCI,~\citet{Tseng_2022} have explored the benefits and tension stemming from applying caring approaches to the provision of security advice and support for IPV survivors. 
Specifically, they show how such approaches can run counter to established routes of measuring the care provided---often quantified and measured against efficiency. 
In response, they recommend to centre users (in their case, IPV survivors) in the evaluation of security care relations. 
Whilst we endorse this recommendation to centre communities in the re-configuration of customer care---we are conscious that currently there is a lack of dialogical structures where these can be co-designed.
We urge the HCI and security and privacy communities to more critically, and profoundly, engage with the implications of neoliberal marketised economies in the shaping of care in the 3DE---specifically in the shaping of what is regarded as appropriate care and how this comes to be measured and modelled (i.e., how the effectiveness of provision is often based on preconceived users and measurable targets formulated without accounting for what happens on the ground). 
We can also more critically engage and reflect on how the very idea of the `\textit{customer}' runs counter to the range of grassroots actions our participants saw as necessary.
Participants wanted to build capacities and confidence to play a role in the collective responsibility to take action, which we discuss next. 

Our findings point to the limited effectiveness of routes focusing solely on security, scams, and fraud ``awareness'' for these communities without considering the broader ecosystem pressures and their preferences for community-led action.  
They show instead the need for \textit{embedded} approaches that can support building grassroots capacities for collectives of action and support, and that can work against the current isolation and fragmentation experienced by participants. 
Existing work in HCI has already moved in this direction, as exemplified in recent work by~\citet{Nicholson_2021}, who attempt to create ``communities of older adults as cyberguardians''. 
Whilst their work focuses on raising awareness using means and languages closer to older adults' everyday life, our findings point more prominently to people and communities creating the conditions for action, knowing that such action can be effective and can lead to change. 
Yet, our findings show how both mindsets (dominant narratives) as well as interpersonal dynamics with our identified snipers play a critical role in shaping confidence to take action. 
Based on this, a promising site of intervention for HCI might be to work to develop critical engagements to support the debunking of dominant narratives that cast fraudsters and scam artists as ``clever'' and by implication those affected as ``thick''.
This could help shift cultural and interpersonal perceptions of shame and embarrassment that our participants articulated.
This would, in turn, create the conditions to build the confidence necessary for communities wanting to take action, thereby reducing both the perceived and actual power of snipers.

Future research may take inspiration from recent work in HCI such as by~\citet{Poon_2023}, which develops a process to support peer-led critical pedagogies for home care workers and explores how such approaches could be re-configured for this domain. 
By adopting critical pedagogies inspired by the work of pioneering activists and scholars, such as~\citet{Freire_2000} and~\citet{hooks_2010}, HCI can support organisations working with communities such as those in our study, develop support that enables the questioning of dominant narratives and systems, as well as enable people and communities to collectively formulate pathways of support to respond to what is critical and important to them. 
At the same time, state institutions that our participants identified as responsible for stopping digital deception, including fraud and scams, must be able to communicate appropriate and effective routes people and communities can take. 
However, as we have argued above, such routes must be designed with the lived experience of people and users in mind through the frame of \textit{positive security}.


\subsubsection{Re-configuring Values:} 
In this paper, we have argued that a marketised 3DE, DPs, and information security models and practice, collectively work together to the benefit of service providers and act as a cutting edge form of neoliberalism~\cite{julier2017}. 
We have explored how interaction and support can be re-configured, but for our participants, this was also about the \textit{values} of the marketised 3DE. 
Effective responses should aim to better engage with the ways different economic models---and their underlining value systems can re-shape how people encounter, feel, and are treated by digital services---and thus mediate relations with service providers, family and friends, and crucially for our participants, against criminal fraudsters.
We urge HCI to engage and question more profoundly the market forces that are at play in the 3DE so as to develop alternatives. 
Practitioners in HCI have begun moving in this direction, for example examining how diverse economies and solidarity economies can inform the design of digital interactions~\cite{Vlachokyriakos_2018, Vlachokyriakos_2017}. 
Yet, notions of diverse economies (see e.g.,~\citet{gibson2013take}), and other economy models (see e.g.~\citet{raworth2017doughnut}) have found limited traction among many HCI and security and privacy researchers (however, see~\citet{Crivellaro2025}). 
Experimenting with how alternative economic models could (re)shape design, exposing DPs~\cite{Mathur_2021}, and re-framing the `customer' or `user' offer challenges to the low-trust and deceptive ecosystems of the marketised 3DE. 

Together, re-configurations across interaction, support, and values are necessary for our participants in northern England to lead lives that they find more fulfilling. 
A short-term focus on computer interaction or improving `customer service' from service providers alone are unlikely to fundamentally rewire the ties and relations of (social) trust that are so integral to the communities we work with.
This is a call for a \textit{positive security} that brings together security and HCI researchers to critically examine what inclusive and equitable forms of security may look like embedded from the grassroots, which we turn to next. 

\subsection{Re-configuring Models for Opportunity}
As much as digital deception has multiple dimensions, as evidenced by the experience of communities in northern England, so then it is required that models must be able to capture such a diversity of use and security needs to challenge snipers' alley. 
Here, we propose that alongside conventional information security and threat models, there must be new forms of modelling that draw on the reconfigurations above, which capture the relational and everyday aspects of security~\cite{Coles-Kemp2017}, and how this relates to an ecosystem of practice in the 3DE.
We propose that in addition to advocating for more diverse and equitable forms of threat model, that threat models sit within a frame of \textit{negative security}. 
Inverting threat to opportunity, we transform the frame and function of HCI and security and privacy research to seek empowerment of people and communities. 
\textit{Opportunity models}, as we term this move, permit an orientation to a frame of positive security whereby users of services are encouraged to identify the positive benefits of digital interaction and develop security strategies to realise those benefits. 
This should address how \textit{re-configurations} across interactions, support, and values must be collectively addressed across the short, medium, and long term.

A reorientation to opportunity seeks to re-configure snipers' alley so that engagement with service providers is viewed as part of the empowerment of people and communities. 
Rather than people simply navigating a (marketised) system that benefits only service providers and protects their systems and data, re-configuring through opportunity must look to relations of trust, and how the marketisation of the 3DE has contributed to a growing distance between providers and people.
As previous work has explored regarding `bottom-up' responses to digital identity from communities~\cite{coles-kemp_digital_2020}, it is essential to understand how information security controls and practices intersect and may reaffirm marketisation.
As people are now often forced through `digital by default' services to enter a constricted digital alley, it is clear from our embedded research that people and communities experience the 3DE differently.
Rather than exploring the protection from threats---from the perspective of enhancing information security---we advocate for a different kind of engagement from a relational and ontologically \textit{positive security} frame. 

An opportunity model asks what can be done to support an individual's ontological security as well as community capability; whether that be to purchase a gift online, to accessing social welfare payments, or having a community forum on Facebook. 
Rather than from the position of what the threats are to these activities, one starts from the perspective of what opportunities from digital connectivity are enabled. 
This challenges a marketised approached where the benefits of digitisation are accrued---typically efficiencies of scale and speed of delivery---by a limited number of actors (e.g., governments and service providers) to one where benefits are more evenly distributed across society~\footnote{For example, in January 2025, the UK Government announced a new drive for use of `artificial intelligence' technologies to enhance economic growth, with little detail on how people and communities' needs may be met over those of the government~\cite{uk_department_for_science_innovation_and_technology_ai_2025}.}.

\section{Conclusion}
This paper has outlined how people and communities in northern England experience a marketised `digital by default' digital economy (3DE). 
The marketisation of service provision from governments and businesses has moved services online and increasingly removed these from communities, and  in the process, made them feel distant and where relations are increasingly `flattened'. 
To manage the increasing distance away from service users, and to mediate digital interactions, information security and threat models have sought to protect systems and data from criminal fraudsters, with an assumption that these are also in the best interest of service users. 
This has occurred concurrently with moves by some service providers to influence users through dark patterns (DPs), leading to individuals to make decisions that might not be in their best interest.
Collectively, our study shows that people, especially with limited capabilities and resources, must navigate blurred lines between `legitimate' and `illegitimate' actors and can experience interactions under systemic stress. 

Collectively, for our study's participants, service providers' digital services have reduced social forms of trust and increased susceptibility to scams and fraud.
We have demonstrated that people must today navigate through a constricted digital alley---which we term ``snipers' alley''---where actors can transfigure to snipers according to service users' capabilities and resources.
Exposure to snipers can negatively affect their everyday lives. 
Whereas some people may not experience the actors---Governments, Business, Criminal Fraudsters, and Friends and Family---as snipers, those that do can experience the entire 3DE through adversarial relations: leading not only to their exclusion from the 3DE but broader society.
It in these collective ecosystem-level understandings of the 3DE that we offer \textit{re-configurations} of digital services.
First, re-configuring  \textit{interaction}, second, re-configuring \textit{support}, and third, re-configuring \textit{values} that sit at the heart of the 3DE. Collectively, interaction, support, and values in the marketied 3DE privilege market dynamics and the proliferation of DPs, rather than on people and communities' needs.
We advocate for inverting information security and threat models that are often designed with a \textit{negative security} frame to \textit{opportunity models} which centre grassroots and embedded perspectives in the frame of \textit{positive security}, so that people can live more fulfilling lives on their own terms.

\begin{acks}
    We would like to Karen Noble, Pallion Action Group and, in particular, the \textit{Pallion Fraud Squad} for their participation and generous contributions to this study. Dwyer's, Heath's and Coles-Kemp's study contributions were funded by the UK EPSRC `Equitable Privacy' project (EP/W025361/1) and the work on the pre-study was funded by Coles-Kemp's `Everyday Safety-Security for Everyday Services' fellowship (EP/N02561X/1).
\end{acks}

\newpage
\bibliographystyle{ACM-Reference-Format}
\bibliography{biblography}

\end{document}